\documentclass[aps,twocolumn,superscriptaddress,footinbib,nobalancelastpage,prl]{revtex4-2}
\usepackage{amsmath,amssymb,dsfont}
\usepackage[colorlinks=true,citecolor=blue]{hyperref}
\usepackage{graphicx}
\usepackage{braket}
\usepackage[table,xcdraw,dvipsnames]{xcolor}

\usepackage{glossaries}
\setacronymstyle{long-short}
\glsdisablehyper
\newacronym{lls}{LLS}{long-lived states}
\newacronym{eth}{ETH}{Eigenstate Thermalization Hypothesis}

\newcommand*{\id}{\mathds{1}}% I

\newcommand{\mun}{\mu/N}
\newcommand{\mucp}{\mu^{p}_c/N}
\newcommand{\mucd}{\mu^{\rho}_c/N}
\newcommand{\mucpd}{\mu^{p,\rho}_c/N}
\newcommand{\tmax}{t_{\rm max}}
\newcommand{\pt}{weak ergodicity breaking transition}

\newcommand{\Lth}{\mathcal{L}_{\mathrm{th}}}
\newcommand{\nth}{N_{\mathrm{th}}}

\begin{document}

\author{Aydin Deger}
\affiliation{Department of Physics and Astronomy, University College London, London WC1E 6BT, United Kingdom}
\affiliation{Interdisciplinary Centre for Mathematical Modelling and Department of Mathematical Sciences, Loughborough University, Loughborough, Leicestershire LE11 3TU, United Kingdom}

\author{Achilleas Lazarides}
\affiliation{Interdisciplinary Centre for Mathematical Modelling and Department of Mathematical Sciences, Loughborough University, Loughborough, Leicestershire LE11 3TU, United Kingdom}

\title{Weak ergodicity breaking transition in randomly constrained model}

\begin{abstract}

Experiments in Rydberg atoms have recently found unusually slow decay from a small number of special initial states. We investigate the robustness of such long--lived states (LLS) by studying an ensemble of locally constrained random systems with tunable range $\mu$. Upon varying $\mu$, we find a transition between a thermal and a weakly non-ergodic (supporting a finite number of LLS) phases. Furthermore, we demonstrate that the LLS observed in the experiments disappear upon the addition of small perturbations so that the transition reported here is distinct from known ones. We then show that the LLS dynamics explores only part of the accessible Hilbert space, thus corresponding to localisation in Hilbert space.

\end{abstract}

\maketitle

{\bf \em Introduction:--}
Isolated quantum systems thermalise: the expectation values of local operators at long times are determined by the values of a small number of conserved quantities (typically, the energy, so that the expectation values coincide with those in the microcanonical ensemble)~\cite{vonNeumann2010}. This is encapsulated in the \gls{eth}~\cite{Deutsch:1991ju,Srednicki:1994dl}, which plays the role for quantum systems that the ergodic hypothesis does for classical, forming the bridge between unitary quantum dynamics and statistical mechanics.

Generic systems satisfy ETH as a matter of course~\cite{DAlessio2015}. Exceptions robust to weak perturbations include many-body localised systems in the presence of disorder~\cite{Mirlin:1996ci,Basko2003,Nandkishore2014} (although there is currently debate on whether these are localised or glassy~\cite{Suntajs2019a,Sierant2020, Sels2021,Abanin2021,Sierant2022}) or quasiperiodic potentials~\cite{Schreiber2015,Iyer:2013gy}. Such systems disobey \gls{eth} either throughout the spectrum~\cite{Huse:2014co} or in a finite fraction of it~\cite{Luitz:2015iv}.

Recently, experiments in Rydberg atoms have observed that certain initial conditions result in abnormally slow decay of the initial state~\cite{Bernien2017}. Consequently, these systems can exhibit a ``weak" violation of ergodicity, meaning that a limited number of non-thermalising eigenstates are present within an otherwise ergodic (thermal) system. This has triggered significant theoretical activity focussing on a class of constrained models, central among them the so-called PXP model~\cite{lesanovsky2011a,Turner2018,Turner2018a,Bull2020a,Bull2019a,Bhilahari2023,Berislav2023}. 

In the PXP model, the bulk of the eigenstates satisfies the \gls{eth}, but a small number (a vanishing fraction of Hilbert space) violate it--these are called scarred states by analogy to the scarred states discussed in quantum chaos \cite{Heller1984}. At the same time they have high overlap with certain experimentally relevant states. Thus, while generic initial states result in thermalisation, starting from one of these few initial states results in the observed anomalous behaviour. The central feature of the PXP model explaining this behaviour is then prove the existence of these scarred states. 

A pertinent question that arises is the stability of these states when subjected to perturbations. Stability with respect to certain local perturbations has been studied in: Ref.~\cite{Khemani2019a}, which found evidence for proximity of the PXP model to an (unknown) integrable model; Ref.~\cite{Surace2020a} which found that the scarred states are unstable (hybridise with the thermal states) in the thermodynamic limit; \cite{Lin2020} which found that a subset of the scars do remain parametrically stable in the thermodynamic limit. This latter model also studies a translationally-invariant modification of the PXP model in which the constraints all have the same, tunable range of range $\alpha$ and which hosts a few low-entropy states. The slow decay of the special initial states of PXP disappears, however, once the spatial range of the constraints is increased beyond that of the PXP model~\cite{Desaules2022}. Finally, Ref.~\cite{surace2022} studies an ensemble of random Hamiltonians, defined as adjacency matrices of random graphs~\cite{Roy2020a}. While the members of this ensemble only have nonvanishing matrix elements between states differing by a single spin flip, a generic member of the ensemble cannot be written as a sum of local terms--the model is therefore intrinsically nonlocal. Additionally, this work focusses on spectral, rather than dynamical, properties. 

In this Letter, we focus on the existence of initial states exhibiting slow decay in models with local PXP-like constraints but of spatially random range. We refer to these states as long-lived states (LLS). We find a phase transition between a  fully ergodic (thermal) phase and one with weakly broken ergodicity supporting \gls{lls} as the constraint strength $\mun$ increases above a threshold. The \gls{lls} exhibit robust oscillations, returning close to their initial states repeatedly before ultimately decaying. These states are not connected to the \gls{lls} present in the clean PXP model: The latter disappear when we introduce local perturbations to the PXP model, and our \gls{lls} only appear once we increase the mean random constraint length.  Meanwhile the bulk spectral properties of the model (such as level statistics) are insensitive to this transition, which is however marked by abrupt changes in both the probability and density of LLS. We finally establish that the \gls{lls} in our model are nontrivial, exploring only a small fraction of the accessible Hilbert space. 

Our work introduces a new class of randomly-constrained models that exhibit a distinct form of weak ergodicity breaking in quantum systems. The key novelty of our work lies in the fact that the weak ergodicity breaking we observe is not a consequence of fine-tuning, as in the case of the PXP model, but rather emerges robustly due to randomness in kinetic constraints and in the absence of explicit disorder. By demonstrating a dynamical phase transition between a fully ergodic phase and one with weakly broken ergodicity, we establish the existence of distinct phases of weak ergodicity breaking with qualitatively different properties and stability. This finding is significant as it provides numerical evidence of a phase transition in the space of constrained quantum many-body systems. Such transitions may be relevant to recent theoretical discussions about the possibility of scarring phase transition in quantum many-body dynamics \cite{Buca2023}.

This paper is organised as follows. We first introduce the model and the class of states we are interested in, then demonstrate the existence of a phase transition between a fully thermal and a weakly non-ergodic phase. This constitutes our main result. We then establish that the \gls{lls} explore only a fraction of the accessible Hilbert space, and finally show that the PXP model is an exceptional member of the ensemble we consider--members of the ensemble with the same constraint range as the PXP model but no translational invariance do not support \gls{lls}.

{\bf \em Model:--} Our randomly-constrained model is described by the Hamiltonian
\begin{equation}
  H=\sum_{i=1}^{N} X_i \prod_{j=1}^{r_i} P_{i-j} P_{i+j},
  \label{eq:H}
\end{equation}
where $X_i, Z_i$ are the usual Pauli spin operators and $P_i=(\id-Z_i)/2$ projects to the down (facilitating) state of spin $i$. Thus, the spin at $i$ can flip only if the $2r_i$ spins a distance $r_i$ on either side are in the facilitating state. We select the $r_i$ independently by drawing random integers from a uniform distribution from interval $[\mu-\epsilon,\mu+\epsilon]$ with a mean $\mu$.

For the model of Eq.~\ref{eq:H}, like for PXP~\cite{Turner2017} as well as models displaying Hilbert space shattering~\cite{Sala2020,Pai2019,Khemani2019c}, Fock space breaks up into disconnected components--spin configurations belonging to one are not reachable from the others by repeated action of the Hamiltonian~\footnote{Viewing the spin model as a single-particle hopping problem on a graph the adjacency matrix of which is the Hamiltonian, not all sites are reachable by allowed hops from all others.}. In this and what follows, we focus on the largest such component of the graph~\footnote{See Supp.~Mat.~for details.}. This sector is always ergodic as far as level statistics and eigenstate properties are concerned~\footnote{See Supp.~Mat.~for details.} but, as we will show, depending on $\mu$, there are non-ergodic states.

Our analysis is primarily focused on the return probability, $\mathcal{L}(t)=\left|\bra{\alpha}\exp(-iHt)\ket{\alpha}\right|^2$ starting from product states $\ket{\alpha}$. We aim to pinpoint those $\ket{\alpha}$ states that exhibit revivals, where the system periodically reverts to a state proximate to its initial configuration. These states will be referred to as \gls{lls}, bearing conceptual resemblance to the two $\mathbb{Z}_2$ states in the PXP model, which have high overlaps with scarred eigenstates.

In what follows we establish that both the probability that such states exist, $p$, and their density $\rho=N_{\mathrm{LLS}}/\mathcal{D_{\mathcal{H}}}$ (with $N_{\mathrm{LLS}}$ the number of such states and $\mathcal{D_{\mathcal{H}}}$ the dimension of the largest connected component of the Hilbert space) departs from $0$ at finite values $\mu^{p,\rho}_c/N$ with $N$ the number of spins; for $\mu<\mu_c^p$ there are no long-lived states, while for $\mu>\mu_c^\rho$ a finite fraction of Fock states result in long-lived oscillations. Within our numerical analysis, $\mu^{p,\rho}$, appear to be either identical or very close. 

At first sight, this appears to contradict known results, since the PXP model is a particular realisation of our model for $\mu=1, \epsilon=0$ but is known to have \gls{lls}. However, we will later show that there is no contradiction: local perturbations in the PXP model eliminate the scarred states, and consequently its \gls{lls}. The \gls{lls} we study only appear in the presence of stronger perturbations. Thus our results indicate the presence of a distinct phase with weakly broken ergodicity, unconnected to the one for the PXP model.

{\bf \em Weak ergodicity breaking transition:--} We use the scaled mean constraining range $\mun$ as a tuning parameter, varying which (for fixed $\epsilon=1$) the probability and density of \gls{lls} departs from 0 (that is, \gls{lls} appear) at some critical $0.2 \leq \mucpd \leq 0.3$. We call this transition {\it \pt} because it is not visible in the usual ergodicity measures such as level statistics or eigenstate properties~\footnote{See Supp.~Mat.~for details.} but rather is only visible in dynamics starting from a small number of initial states.

To be more concrete, we first provide a precise definition of the \gls{lls} and then characterise these states in terms of their strength and persistence. Starting with a given Fock state $\ket{\alpha}$ we evolve it using Eq.~\ref{eq:H} up to some time $\tmax$ and then calculate the return probability $\mathcal{L}(t)$. For $\alpha$ to qualify as a \gls{lls}, we count the number of times, $\nth$, that the return probability $\mathcal{L}(t)$ goes above a given threshold $\Lth$. In what follows we define an \gls{lls} as one for which $\nth\geq 3$ for $\Lth=0.5$. We have checked that our main findings are qualitatively the same for other definitions of $\nth$ and $\Lth$~\footnote{See Supp.~Mat.~for details.}. 

Under this definition, the $\mathbb{Z}_2$ states are categorised as \gls{lls} in the PXP model. Consequently, this definition facilitates the connections between \gls{lls} and non-thermal many-body states. That is, the presence of \gls{lls} for a specified $\mu$ implies the existence of eigenvectors that highly overlap with the \gls{lls}, giving rise to quantum many-body scarring.

\begin{figure}[!t]
  \includegraphics[width=\linewidth]{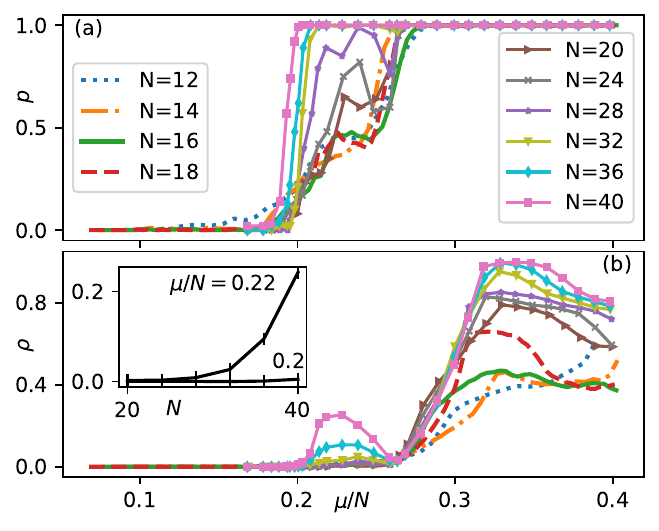}
  \caption{Weak ergodicity transition. (a) Probability of finding at least one LLS for mean constraint range $\mun$; for $N<18$ we averaged over 1000 realisations per $\mun$ and up to time $\tmax=18$, while for $N>20$ we used 100 realisations up to $\tmax=50$. (b) Density $\rho$ of LLS versus of $\mun$ jumps sharply away from 0 around $\mun \sim 0.2$. The inset  shows $\rho$ vs system size $N$ for $\mun=\{0.2,0.22\}$; above $\mun\sim 0.2$,  $\rho$ increases with system size. We believe the bump between $\mun=0.2$ and $0.3$ to be a finite-size effect (see text). This indicates a phase transition in the thermodynamic limit.}
  \label{fig:density}
\end{figure}

In Fig.~\ref{fig:density}, we report conclusive evidence of the aforementioned phase transition. The top panel (a) shows \gls{lls} probability for different system sizes as a function of $\mun$ while the bottom panel (b) shows the density of \gls{lls}--the fraction of Fock states in the largest connected cluster that are \gls{lls}. Both of these measures depart from 0 at around $\mun\approx 0.2$. 

For the probability $p$, the trend in Fig.~\ref{fig:density}(a) with increasing system size makes it clear that, in the thermodynamic limit, $p$ rapidly rises to 1 above $\mun=\mucp\approx 0.2$, so that at least one \gls{lls} appears. Meanwhile, the density $\rho$ displays a more intricate behaviour as depicted in Fig.~\ref{fig:density}(b). Initially, it too departs from 0 at $\mun\approx\mucd$ and just above it trends to a finite value as $N$ increases. At $\mun>\mucd\approx 0.2$, the density increases with system size (see inset), so that one can confidently state that for $\mun>\mucd$ a finite density of Fock states are \gls{lls}. While there exists a parameter regime $0.2<\mun<0.3$ where the behaviour appears to be non-monotonic, with two peaks appearing, we believe this to be a finite-size effect for the following reason. In the Supp.~Mat.~we show plots of both $p$ and $\rho$ for different thresholds $\Lth=0.6$ and $0.7$, larger than the value $0.5$ used here. We notice that for $\Lth=0.6$, the density $\rho$ also picks up this peak for higher values of $N$. For the probability $p$ at $L_{th}=0.5$ the trough gets filled in above $N\geq 30$ whereas for $L_{th}=0.6$ at $N\geq 40$, so we expect that the trend continues at $L_{th}=0.7$ with the trough getting filled in at a larger $N$ which we cannot access ~\footnote{See Supp.~Mat.~for details.}. This suggests again that this is not a few-body effect, as it only appears at large-enough sizes. We have not been able to ascertain the origin of this behaviour, so that it remains to be determined in future work. Since $p$ is more sensitive than $\rho$ (as it detects a single \gls{lls}), we conclude that this behaviour will also be mirrored for $\rho$, but at larger (and inaccessible to us) sizes, and thus believe the actual transition in $\rho$ to lie at the point where it departs 0 for the first time, around 0.2. In conclusion, from our results in Fig.~\ref{fig:density} it is evident that, first, \gls{lls} definitely appear for $\mun>\mucp\approx 0.2$ and, second, that a \emph{finite density} of such states appears for $\mun>\mucd\approx 0.2$.

Two obvious questions present themselves at this point. First: Could it be that the largest connected component in Hilbert space is small enough that we are simply seeing recurrences because of its finiteness (as opposed to the revivals being due to the dynamics exploring only a subspace of that)? Second: From Fig.~\ref{fig:density}, it would appear that the PXP model, a specific realisation of $\mun=1/N$, should display no \gls{lls}. But, as is well known, it does have \gls{lls}; so how can our results be reconciled with that? 

We answer each of these questions in turn.

{\bf \em Truncated Lanczos Iterations:--} In order to address the first question, we analyse the fraction of the Hilbert space of the largest connected component explored by the dynamics starting from the \gls{lls}. To do so we use the Lanczos algorithm. In brief, this involves the following steps: Given an initial vector $\ket{\alpha}_0$ and a matrix $H$, at the n$^{th}$ step one constructs a vector $\ket{\beta_n}=H\ket{\alpha_n}-u_n\ket{\alpha_n}$ with $u_n=\bra{\alpha_n}H\ket{\alpha_n}$ and $v_n=\sqrt{\left<\beta_n\right|\left.\beta_n\right>}$; then $\ket{\alpha_{n+1}}=\ket{\beta_{n}}/v_{n+1}$. After $m$ such iterations, one forms the matrix $H_{\rm eff}(m)=VTV^\dagger$ where $V$ has the $\ket{\alpha_n}$ as columns and $T$ has the $u_n$ on the main diagonal and the $v_n$ on the first off-diagonal constitutes an approximation to $H$. In principle, $m=\mathcal{D}_H$ exactly reproduces $H$. 

Our approach \emph{truncates} this procedure at some order $m$, determined by minimising the following cost function with respect to $m$:
\begin{equation}
  I = \frac{1}{\tmax} \min_{m}\int_0^{\tmax}dt \big| \mathcal{L}(t) - \mathcal{L}_{\rm TLI}(m,t) \big|.
  \label{eq:cost}
\end{equation}
Here, $\mathcal{L}(t)$ represents the return probability of a \gls{lls} evolved with \eqref{eq:H}, while $\mathcal{L}_{\rm TLI}(m,t)$ denotes the return probability with an effective Hamiltonian $H_{\rm eff}(m)$ constructed by using the Lanczos algorithm. We terminate the minimization procedure when $I \leq 0.01$. The aim of this truncation is to determine what fraction of Hilbert space is explored by the dynamics by explicitly constructing it--its dimension is clearly $m_c$. A similar approach has recently been used to differentiate between localised and chaotic quantum systems \cite{alaoui2023method}.

\begin{figure}[!t]
  \includegraphics[width=\linewidth]{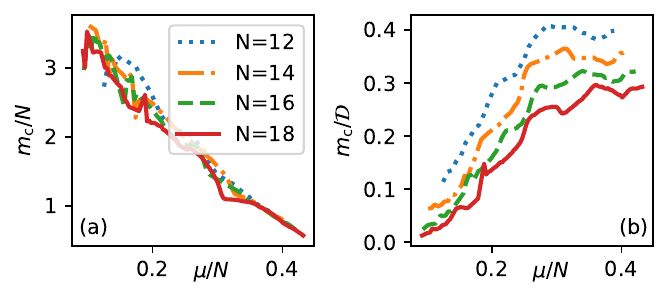}
  \caption{Minimum number of states $m_c$ required to reproduce the return probability of \gls{lls} for a given constraint range $\mun$. (a) Scaled by system size $N$; $m_c/N$ does not depend on system size, while (b) $m_c/D_H$ decreases with system size, indicating that the \emph{fraction} of Hilbert space that is involved in the dynamics decreases with system size. In both panels, averaging is performed over 1000  realisations for each $\mu$.}
  \label{fig:lanczos}
\end{figure}

For each realisation at a given $\mun$, we obtain the $m_c$ for each \gls{lls}, then average over all the \gls{lls} and realisations. The resulting $m_c$ as a function of $\mun$ is shown in Fig.~\ref{fig:lanczos}, scaled by $N$ (left panel) or $\mathcal{D}_H$ (right panel). Our findings elucidate several key aspects. Firstly, the number of states $m_c$ required is linear in system size $N$ for given $\mun$ (left panel), exhibiting a universal behaviour. Secondly, a notable decrease in $m_c$ is observed with increasing constraint range (left panel). Lastly, the \emph{fraction} of the Hilbert space involved in the dynamics for given $\mun$ decreases with increasing system size (right panel). 

Let us summarise this calculation and the conclusions to be drawn from it. We have shown that the dynamics of the \gls{lls} is restricted to a Krylov subspace of a dimensionality that is a decreasing fraction of the dynamically accessible Hilbert space. This implies that the \gls{lls} are caused by nontrivial dynamics inside the largest connected cluster, rather than simply a result of the dimension of the largest cluster decreasing with $\mun$.

\begin{figure}[!t]
  \includegraphics[width=\linewidth]{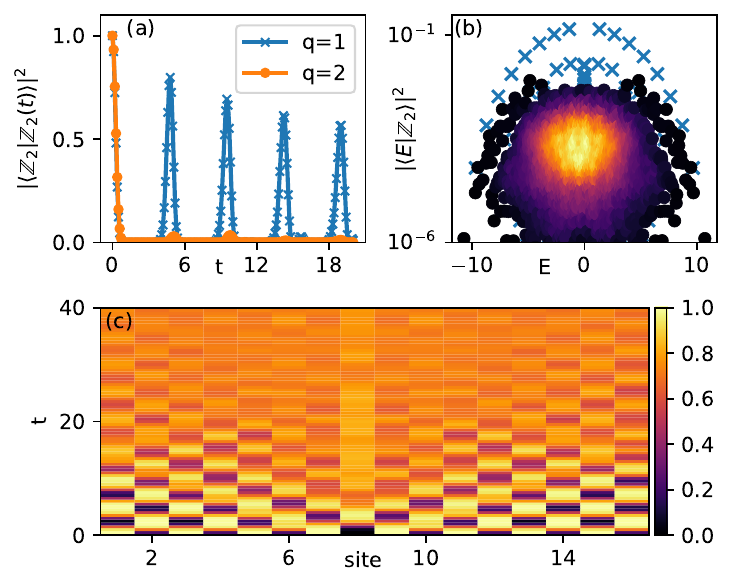}
  \caption{Scars vanish due to a defect in the constrained range; we set $r_{i_0}=q$ for a single site $i_0$, and leave $r_i=1$ for $r_i\neq i_0$. (a) Return probability starting from $\mathbb{Z}_2$ state. $q=1$ (blue, point) corresponds to the standard PXP model whereas $q=2$ (orange, square) means that there's a single defect with strength equal to 2. (b) Overlap of the $\mathbb{Z}_2$ state with the eigenstates. (c) Spatiotemporal profile of density starting from $\mathbb{Z}_2$ state with a single defect $q=2$.}
  \label{fig:impurity}
\end{figure}

{\bf \em Disappearance and re-emergence of quantum scars:--} We now come to the apparent contradiction mentioned earlier, namely, that in Fig.~\ref{fig:density} the probability for \gls{lls} to exist for $\mun=1$ vanishes, while the PXP model is a particular realisation of $\mun=1/N\rightarrow_{N\rightarrow\infty} 0$. The resolution of this paradox is that the PXP model is a singular point: Changing a single $r_i\neq 1$ results in a rapid decay of the oscillations and destroys the unique spectral structure. Fig.~\ref{fig:impurity}(a) shows the return probability starting from a $\mathbb{Z}_2$ state for both standard PXP and for the PXP model modified by setting a \emph{single} $r_{i_0}=2$ for some $i_0$. In Fig.~\ref{fig:impurity}(b), we show the overlap of the $\mathbb{Z}_2$ state with the eigenstates of the model for the case of the standard PXP (blue) and our perturbed model with $r_{i_0}=2$; the characteristic peaks that are known from the PXP model disappear (same behaviour is observed for all other fock states in the largest sector). Thus, weak, local perturbations of the PXP model disrupt the scars (thus also the \gls{lls}, $\mathbb{Z}_2$ and $\mathbb{Z}_2'$ in the notation of~\cite{Turner2017}), which implies that the phase with \gls{lls} that we uncover at higher $\mun$ is not connected to the scarred phase of the PXP model.

At this point, it's natural to question how a \emph{local} perturbation can disrupt \emph{global} quantities such as the eigenstate overlaps; after all, in the thermodynamic limit, a local perturbation should be negligible, so how is it capable of eliminating the oscillations? The resolution to this paradox is that the eigenstates most directly pertain to infinite-time results of observables via the eigenstate thermalisation hypothesis (ETH). Thus, as we show in Fig.~\ref{fig:impurity}(c), the oscillations starting from a Neel $\mathbb{Z}_2$ state with a model with a single $r_{i_0}=2$ results in oscillations decaying inside a lightcone spreading out from $i_0$; only after a time $\propto N$ will they decay everywhere. The effect is visible in spectral properties such as the eigenstate expectation values only because those are relevant for the infinite-time limit.

{\bf \em Conclusion:--} In this work we have studied an ensemble of random, local, constrained models parameterised by the mean constraint range $\mu$. We find that typical members of the ensemble transition from a low-$\mu/N$ phase with no \gls{lls} to a high-$\mu/N$ phase with high density of \gls{lls}. This appears to contradict known results for the PXP model, which is a special case for $\mu=1$. We reconcile the two results by showing that increasing the constraint range at a single sit of PXP causes its unique spectral features to disappear. 

There remain a number of open questions on the nature and origins of these \gls{lls} remain. Numerical experiments with the Lanczos methods (not shown) suggest that the dynamics of some, but not all, of our \gls{lls} is well-reproduced by replacing the Hamiltonian by the adjacency matrix of a hypercube with the initial \gls{lls} as a node~\cite{Turner2018a}. Can this idea be extended to include all of them? Do all \gls{lls} correspond to dynamics on a small number of special subgraphs (analogously to how the PXP \gls{lls} are due to adjacency matrices corresponding to a hypercube, or the ``motifs'' of Ref.~\cite{surace2022})? We leave the answers to such questions for future work.

\begin{acknowledgments}
We thank Juan P. Garrahan for helpful discussions. This work was supported by EPSRC Grant No. EP/V012177/1.
\end{acknowledgments}

\appendix

\subsection{Determination of connected components}

Here, we describe the construction of the Hamiltonian of our model. This can be equivalently considered as finding the largest connected sector, where each Fock state fulfils the imposed kinetic constraints. For a translationally invariant constraining list, the Hamiltonian can be constructed on a momentum basis, e.g., for the PXP model. However, since the constraints in our model are randomly selected at each site, we cannot expect it to be translationally invariant. Consequently, we employ the following steps to identify the largest sector in the Hilbert space for our randomly constrained model.

To begin, we start with a fully facilitating state, denoted as $|f\rangle=| 00\ldots000 \rangle$, and find a list of neighbouring states, $|n_i\rangle$, that are one Hamming distance away from $|f\rangle$. The Hamming distance between two states can be determined simply by applying the XOR operation on their respective bits. Next, we eliminate the states that do not satisfy the kinetic constraint conditions. For instance, a spin at site $j$ can flip if all spins between $j - r[j]$ and $j + r[j]$ apart from itself, i.e., the spin at $|n[j]\rangle$ is facilitating. Subsequently, we repeat the same algorithm to find all neighbours of $|n_i\rangle$, continuing this iteration until we have constructed a complete set of eligible Fock states that adhere to the given kinetic constraints. This set corresponds to the basis states of our Hamiltonian.

\begin{figure}
  \includegraphics[width=1\linewidth]{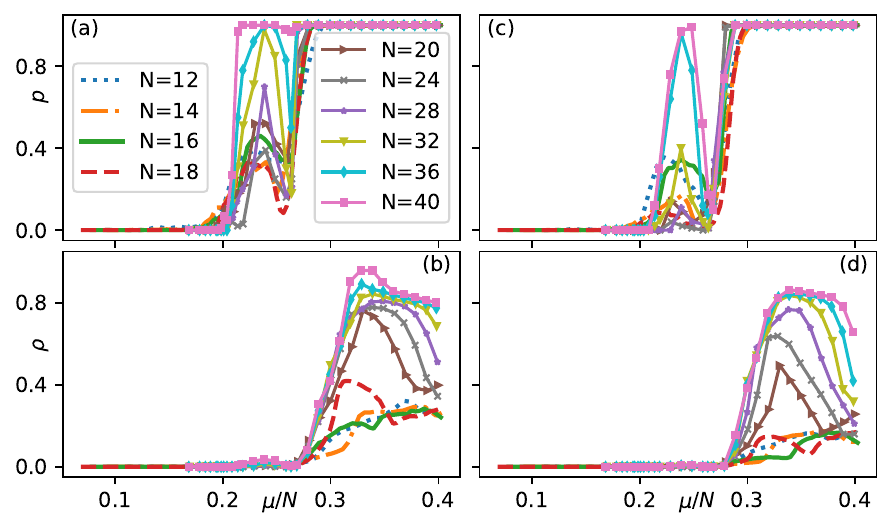}
  \caption{Probability and density for threshold=$0.6$ (a,b) and threshold=$0.7$ (c,d).}
  \label{fig:suppPRho}
\end{figure}

\subsection{Different LLS threshold and finite-size effect}

Here we demonstrate how our main result would appear if we use different threshold values, namely the number of times $\nth$ the return probability exceeds the threshold $\Lth$. In the main text, we only considered $\nth\ge3$ and $\Lth=0.5$. In Fig.~\ref{fig:suppPRho}, we present the probability $p$ and density $\rho$ for LLS with threshold values $\Lth={0.6,0.7}$. As can be observed, qualitatively, they do not differ from the main results. We observe that increasing $\Lth$ suppresses the distinct peak occurring around $\mun\sim0.2$ for the sizes considered here. We anticipate that for larger sizes, the peak will reappear, and the probability and density will reach a finite value for $\mun > 0.2$. Additionally, we note that modifying the definition of LLS slightly shifts the critical point towards larger values of $\mun$. This outcome is expected since in the other limit, i.e., $\Lth\ll0.5$, the critical point would trivially occur at $\mun=0$.

\begin{figure}
  \includegraphics[width=1\linewidth]{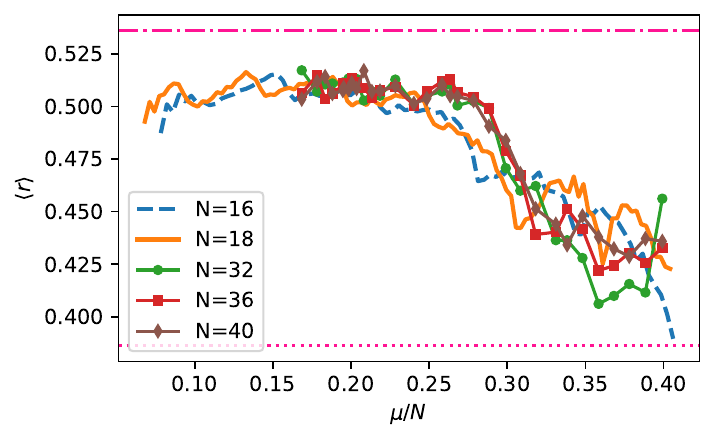}
  \caption{Level statistics.}
  \label{fig:suppLevel}
\end{figure}

\subsection{Level statistics}

The level statistics can be determined by evaluating the spacing between energy levels. Here we use the standard formula:
\begin{equation}
r_n = \min(\delta_n,\delta_{n+1})/\max(\delta_n,\delta_{n+1}),
\end{equation}
where $\delta_n$ is the gap between the $n$th and $\{n-1\}$th energy eigenvalues. The statistical properties of energy eigenvalues have been observed to provide crucial insights into many-body localisation (MBL), quantum chaos, and the symmetries present in the system. In particular, in the case of repulsion between eigenvalues, the mean level statistics yield $\langle r_n \rangle \approx 0.536$ which corresponds to the Gaussian orthogonal ensemble. The Poisson-level statistics, $\langle r_n \rangle \approx 0.386$, emerge when there is no correlation between eigenvalues within the system. This diagnostic is commonly used to identify strong ergodicity-breaking transitions, e.g. MBL transition.

In Fig.~\ref{fig:suppLevel}, we show the level statistics averaged over each realization at $\mun$ data points.Note that we focus on level statistics only in the largest sector. We observe that for $0.25 < \mun < 0.3$, the mean level statistics display a decline, followed by a quick levelling off above the Poisson line for larger $\mun$ values. This suggests that while the energy spectrum undergoes quantitative changes near the critical point, no significant behaviour can be inferred from the drop as the system size increases. Consequently, this standard diagnostic does not allow for the investigation of the \emph{\pt} described in the main text.

%\bibliography{biblio}


\end{document}